\documentclass{PoS}

\usepackage{psfrag}

\title{Inverse magnetic catalysis in QCD}

\ShortTitle{Inverse magnetic catalysis in QCD}

\author{Falk Bruckmann\thanks{Supported by the the Alexander von Humboldt
    foundation} \\ 
    Institute for Theoretical Physics, Universit\"at Regensburg \\ 
    D-93040 Regensburg, Germany \\ 
    E-mail: \email{falk.bruckmann@physik.uni-regensburg.de}}

\author{Gergely Endr\H{o}di\thanks{Supported by the EU under grant number ITN
    STRONGnet 238353.} \\ 
  Institute for Theoretical Physics, Universit\"at Regensburg \\ 
  D-93040 Regensburg, Germany \\ 
  E-mail: \email{gergely.endrodi@physik.uni-regensburg.de}}

\author{\speaker{Tamas G. Kovacs}\thanks{Supported by the Hungarian
    Academy of Sciences under ``Lend\"ulet'' grant No.\ LP2011-011 and by the
    EU Grant (FP7/2007-2013)/ERC No. 208740.} \\ 
   Institute of Nuclear Research of the Hungarian Academy of Sciences,\\ 
   Bem t\'er 18/c, H-4026 Debrecen,  Hungary \\ 
   E-mail: \email{kgt@atomki.mta.hu}}

\abstract{We propose a physical mechanism for inverse magnetic catalysis, the
suppression of the chiral condensate by an external magnetic field in QCD
around the critical temperature. We show that this effect, seen in lattice
simulations, is a result of how the sea quarks react to the magnetic field. We
find that the suppression of the condensate happens because the quark
determinant can suppress low quark modes by ordering the Polyakov loop. This
mechanism is particularly efficient around $T_c$ where the Polyakov loop
effective potential is flat and the determinant can have a significant
ordering effect. Our picture suggests that for the description of QCD in large
magnetic fields it is crucial to properly capture the interaction between the
Polyakov loop and the sea quarks, both in low-energy effective models and on
the lattice.}

\FullConference{31st International Symposium on Lattice Field Theory - LATTICE
  2013\\ July 29 - August 3, 2013\\ Mainz, Germany}

\begin{document}

\section{Introduction}

\begin{figure}
\begin{center}
\includegraphics[width=0.5\columnwidth,keepaspectratio]{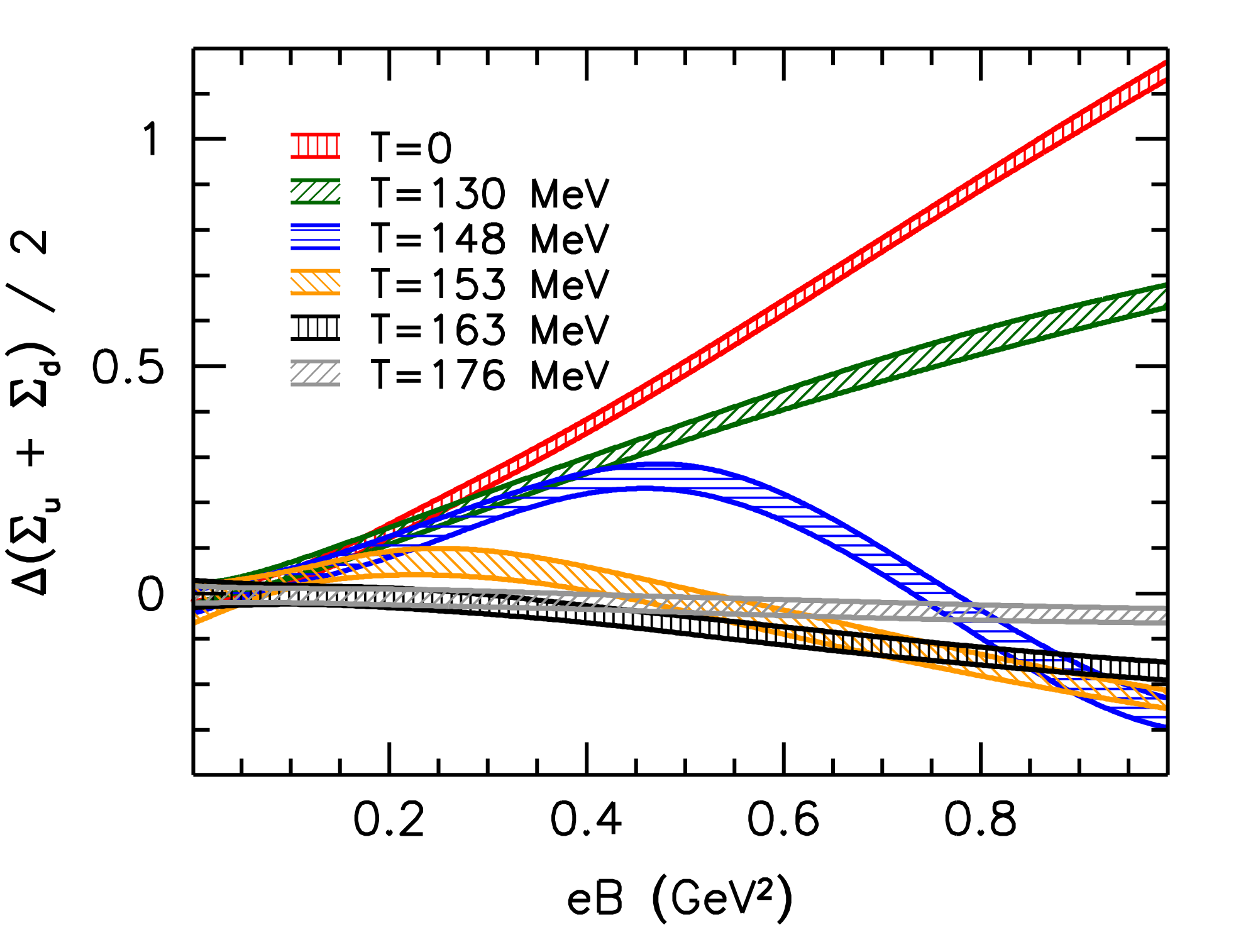}
\vspace{6ex}
\caption{\label{fig:pbp_B_T} The dependence of the quark condensate on the
  magnetic field, computed at different temperatures around $T_c$.}
\end{center}
\end{figure}

In heavy-ion collision experiments rapidly moving ions create large magnetic
fields. There have been speculations how these large magnetic fields can
influence the thermodynamics of these strongly interacting systems. For some
time the common wisdom was that at any temperature the magnetic field should
enhance the chiral condensate, an effect termed ``magnetic catalysis'' (see
\cite{Shovkovy:2012zn} for a review). This scenario was supported by
calculations based on low-energy effective models and also by the first
lattice QCD simulations. However these first lattice studies were based either
on the quenched approximation \cite{lattice_quenched} or they were performed
using heavier than physical quarks and without a proper continuum
extrapolation \cite{D'Elia:2010nq,D'Elia:2011zu}.  Later on, another lattice
study, now using physical quark masses and performing an extrapolation to the
continuum limit, reached a different conclusion \cite{Bali:2011qj}. There it
was found that while at low temperature there is magnetic catalysis, around
the crossover temperature, $T_c$, the reaction of the system to an external
magnetic field is more complicated. The condensate is a non-monotonic function
of the magnetic field and a large enough magnetic field actually suppresses
the condensate rather than enhancing it. To illustrate this point, in
Fig.\ \ref{fig:pbp_B_T} we show the dependence of the quark condensate on the
magnetic field at various temperatures around $T_c$ (figure taken from
Ref.\ \cite{Bali:2012zg}). 

In the present work we offer a physical explanation of why this can happen. We
show that there are two competing mechanisms through which the magnetic field
influences the quark condensate. One involves valence quarks and that enhances
the condensate, the other involves sea quarks and that suppresses the
condensate. The non-monotonic behaviour observed around $T_c$ is a result of
these two competing effects. We also identify the relevant gauge degrees of
freedom playing the most important role in the sea quark suppression
mechanism. These degrees of freedom are the Polyakov loops. We argue that
effective low-energy models can account for this effect only if they properly
take into account the interaction of sea quarks with the Polyakov loop. A more
detailed account of this work can be found in Ref.\ \cite{Bruckmann:2013oba}.

\section{Valence versus sea effect}

In order to understand how the external magnetic field influences the quark
condensate, our starting point is the path integral expression for the
condensate, 
\begin{equation}
  \langle \bar{\psi}\psi \rangle (B) = \frac{1}{Z} \int dA \;
      \mbox{e}^{-S(A)} \; 
      \underbrace{ \det[D(A,B)+m]}_{\mbox{``sea''}} \; 
      \underbrace{ \mbox{Tr} \left[ (D(A,B)+m)^{-1}
          \right]}_{\mbox{``valence''}},
\end{equation}
where the integration is over all the gauge field configurations. For
illustration we use this schematic expression containing only one quark
flavour of mass $m$; the generalisation to several flavours is
straightforward. The dependence of the condensate on the external field, $B$,
comes from two sources. Firstly, the quark determinant in the measure depends
on $B$ which means that changing $B$ will change the relative weight of
different gauge configuration. Secondly, the operator itself is also dependent
on $B$, that is, the spectrum of the Dirac operator in a fixed gauge
background changes with $B$. Using the terminology of
Ref.\ \cite{D'Elia:2011zu} we call the first source of $B$ dependence ``sea''
effect and the second one ``valence'' effect.

\begin{figure}
\begin{center}
\includegraphics[width=0.5\columnwidth,keepaspectratio]{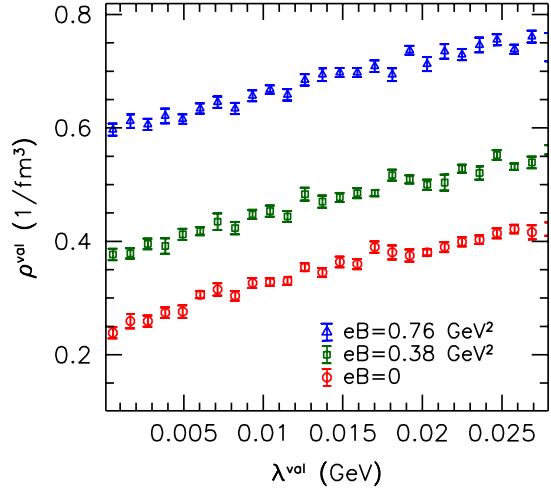}
\caption{\label{fig:specdens} The spectral density of the staggered Dirac
  operator around zero computed with different magnetic fields in the Dirac
  operator. In all three cases the averaging was done over the same set of
  gauge configurations generated with zero magnetic field.}
\end{center}
\end{figure}

Let us first look at the valence effect, how the spectrum of the Dirac
operator in a fixed gauge background reacts when the magnetic field is
switched on. In Figure \ref{fig:specdens} we show the spectral density of the
staggered lattice Dirac operator around zero with three different values of
the magnetic field in the Dirac operator. It is important to note that in all
three cases the average spectral density was computed over the same set of
gauge configurations, generated with zero magnetic field. Thus this figure
demonstrates the valence effect only. We can easily see that the external
magnetic field drives the low-lying Dirac modes closer to zero and thus
enhances the spectral density. According to the Banks-Casher relation, for
vanishing quark mass, the quark condensate is proportional to the spectral
density of the Dirac operator at zero. By continuity we expect that if the
quark mass is small, the condensate is still dominated by the lowest part of
the spectrum. Thus an enhancement of low Dirac modes by the magnetic field
implies and enhancement of the condensate.  This is the basic mechanism behind
magnetic catalysis happening well below $T_c$.

\subsection{Sea quarks and the Polyakov loop}

\begin{figure}
\begin{center}
\includegraphics[width=0.5\columnwidth,keepaspectratio]{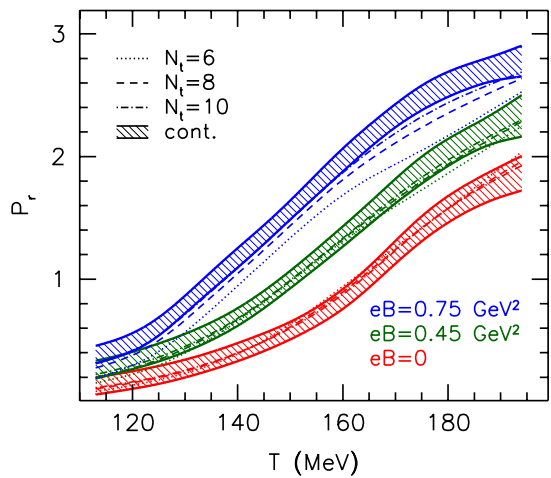}
\caption{\label{fig:ploop_cont} The renormalised average Polyakov loop as a
  function of the temperature for three different values of the background
  magnetic field. The different types of dashed line represent data obtained
  at different lattice spacings and the shaded regions correspond to the
  continuum extrapolations together with their uncertainties.}
\end{center}
\end{figure}

Let us now look at how the magnetic field in the quark determinant influences
the condensate. Switching on the magnetic field will change the relative
weight of the different gauge configurations. To see how this changes the
typical gauge configurations contributing to the path integral we look for
gauge field degrees of freedom that could play an important role in this
mechanism. In Figure \ref{fig:ploop_cont} we plot how the average Polyakov
loop changes across the transition for three different values of the magnetic
field. It is apparent from the figure that around the critical temperature the
magnetic field strongly enhances the Polyakov loop.

To understand why the Polyakov loop is so strongly affected by the magnetic
field we look at the quark action, the logarithm of the determinant,
\begin{equation}
   S_q = -\log \det(D+m) =  -\sum_i
                       \log \left( \lambda_i+m \right),
\end{equation}
where $\lambda_i$ are the eigenvalues of the Dirac operator. For small
quark mass the fluctuations of this action are dominated by the small
eigenvalues of the Dirac operator, so the determinant will strongly suppress
those gauge configurations that have a large number of small Dirac
modes. Since the magnetic field enhances small Dirac eigenvalues, switching on
the magnetic field will amplify the suppression of small Dirac modes by the
quark determinant.

How can we understand the connection of this effect to the Polyakov loop? How
is the ordering of the Polyakov loop connected to the number of small Dirac
modes? To see this we recall that at low temperature, well below the
transition, where the Polyakov loop is disordered, there are typically many
small Dirac modes. This is how chiral symmetry is spontaneously broken
there. In contrast, above $T_c$, chiral symmetry is restored and there are
much less small Dirac modes. The physical reason for this is that above $T_c$
the Polyakov loop is ordered and the lowest quark modes are similar to the
Matsubara modes with the corresponding eigenvalues being roughly proportional
to the temperature. This is the mechanism behind chiral symmetry restoration
above $T_c$. 

\begin{figure}
\begin{center}
\psfrag{action difference}{$\Delta S_q \;\;\; \rightarrow$ {\small exponentially
  suppressed by $B$}} 
\psfrag{Ploop}{$P$}
\includegraphics[width=0.5\columnwidth,keepaspectratio]{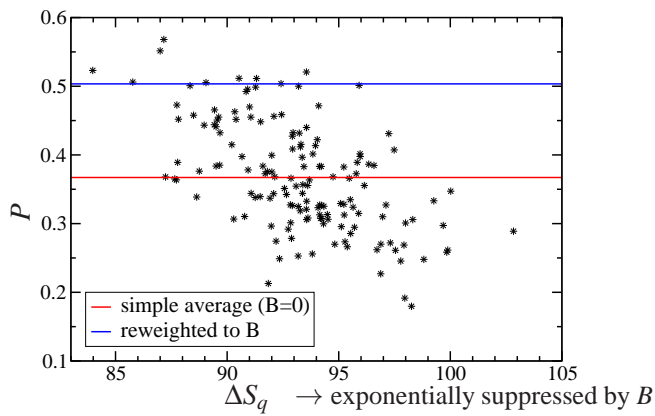}
\caption{\label{fig:plp_vs_ldet} Scatter plot of the Polyakov loop versus
  the change in quark action, $\Delta S_q$, when the magnetic field is
  switched on. Each data point corresponds to a $10^3 \times 4$ gauge
  configuration generated with zero magnetic field close to $T_c$. The lower
  (red) line indicates the simple average of the Polyakov loop for these
  configurations while the upper (blue) line shows the average computed
  by reweighting each term with the corresponding change in the quark
  effective action.}
\end{center}
\end{figure}

In this way the quark action effectively works to order the Polyakov loop and
restore chiral symmetry.  Since the magnetic field enhances small Dirac modes,
this ordering effect of sea quarks is also enhanced by the magnetic
field. This means that configurations that have a small average Polyakov loop
and as a result more small Dirac modes, get suppressed when the magnetic field
is switched on. This is analogous to what happens in the presence of more
quark flavours. Introducing more quark flavours will amplify the ordering
effect of the determinant on the Polyakov loop and thus chiral symmetry is
restored at a lower temperature. The magnetic field has a similar effect by
pushing small Dirac modes down and enhancing the spectral density at the low
end. 

To demonstrate explicitly how this happens, in Fig.\ \ref{fig:plp_vs_ldet} we
show a scatter plot of the Polyakov loop versus the change in quark action
when the magnetic field is switched on,
\begin{equation}
  \Delta S_q = \log \det (D(0,A)+m) - \log \det (D(B,A)+m),
\end{equation}
for a set of gauge configurations generated with zero magnetic field. In the
same figure we also show the simple average of the Polyakov loop over these
gauge configurations (lower, thick red line) and the average computed by
reweighting each configuration using the corresponding change in the quark
action. The reweighted Polyakov loop is clearly much more ordered than the
original one, in accordance with Fig.\ \ref{fig:ploop_cont}.

\section{The Polyakov loop and the condensate}

\begin{figure}
\begin{center}
\psfrag{psibarpsi}{$\bar{\psi}\psi$} 
\psfrag{ploop}{$P$}
\includegraphics[width=0.5\columnwidth,keepaspectratio]{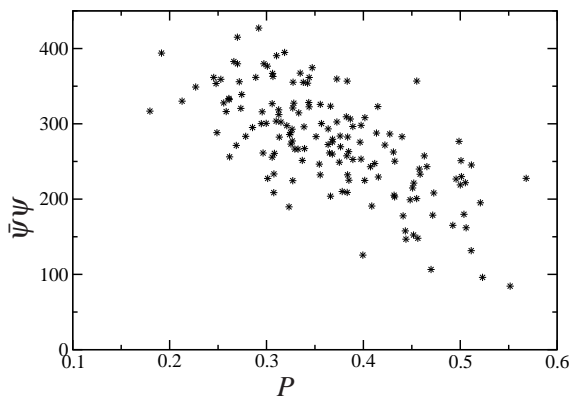}
\caption{\label{fig:plp_vs_pbp} Scatter plot of the The Polyakov loop versus
  the quark condensate. Each data point corresponds to a $10^3 \times 4$ gauge
  configuration generated with zero magnetic field close to $T_c$.}
\end{center}
\end{figure}

We saw that switching on the magnetic field orders the Polyakov loop and that
implies a suppression of small quark modes by the determinant (sea). The very
same small Dirac modes are the key to understand how the quark condensate is
influenced by the magnetic field. To see this, we note that the condensate can
be written as
\begin{equation}
 \bar{\psi}\psi = \mbox{tr} (D+m)^{-1} \approx \sum \frac{1}{\lambda_i+m},
\end{equation}
where the sum is over the whole spectrum of the Dirac operator and $m$ is the
quark mass. For small quark mass this sum is increasingly dominated by the
lowest part of the Dirac spectrum. Thus the suppression of small Dirac modes
by the magnetic field implies a suppression of the quark condensate. 

To illustrate this point, in Fig.\ \ref{fig:plp_vs_pbp} we show a scatter plot
of the quark condensate versus the Polyakov loop on the same set of
configurations that we used for Fig.\ \ref{fig:plp_vs_ldet}. The strong
correlation between the condensate and the Polyakov loop indicates that these
two quantities are both intimately connected to small quark modes as we
described above. 

\section{Competition between ``sea'' and ``valence''}

We saw that when the magnetic field is switched on, sea quarks suppress those
gauge configurations that have many small Dirac modes. At the same time the
magnetic field in the operator enhances the condensate (valence). Both of
these effects depend on how small quark modes react to the external magnetic
field. The non-monotonic behaviour of the condensate with respect to
the magnetic field can be explained by these two competing effects. Around the
critical temperature, if the magnetic field is strong enough, the suppression
of the condensate through the sea quarks wins, resulting in a condensate
decreasing for larger fields.

It is curious why the sea suppression of the condensate can be so effective
exactly around the critical temperature. The scenario that we described can
also explain this. We saw that the most effective way of sea quarks to
suppress small Dirac modes is to order the Polyakov loop. A small contribution
from the magnetic field in the determinant to the Polyakov loop effective
potential is usually not enough to have a significant ordering effect. This is
true everywhere except around the critical temperature. Since at $T_c$ there
is a cross-over, the minimum of the Polyakov loop effective potential there is
very shallow and a small contribution to it from the magnetic field can have a
significant ordering effect. 

\section{Conclusions}

We described the mechanism that is responsible for inverse magnetic catalysis,
the suppression of the quark condensate by an external magnetic field around
$T_c$. We showed that the key to this is the effect of the magnetic field on
small quark modes in the Dirac operator. The magnetic field enhances small
quark modes which normally, through a Banks-Casher type argument, implies an
enhancement of the condensate. This is what we called the valence
effect. However, around $T_c$ the magnetic field in the determinant
significantly alters the typical gauge configurations that contribute to the
path integral. As a result, the Polyakov loop gets more ordered, small Dirac
modes and also the condensate are suppressed. This is what we called sea
effect.

Our picture also suggests the most important criteria that any effective model
or lattice simulation has to fulfil in order to be able to properly account
for inverse magnetic catalysis. Firstly, both in the quark determinant and in
the operator measuring the condensate, the dominance of the lowest part of the
Dirac spectrum is strongly dependent on the quark mass being small. Therefore,
it is essential to use physical quark masses. Secondly, the sea effect depends
on the interaction of the condensate and the Polyakov loop which is
``mediated'' by the small quark modes. Therefore, these degrees of freedom
have to be included in some form in any low energy effective model used to
describe how the magnetic field affects the condensate.

\end{document}